\documentstyle[12pt]{article}\textheight 230mm\textwidth 150mm
\newcommand{\be}{\begin{eqnarray}}
\newcommand{\ee}{\end{eqnarray}}
\newcommand{\eel}[1]{\label{#1}\end{eqnarray}}
\newcommand{\r}[1]{(\ref{e:#1})}

\newcommand{\vb}{{\cal h}}
\newcommand{\hb}{{\cal i}}

\newcommand{\Lra}{{\Leftrightarrow}}

\newcommand{\nn}{\nonumber}

\newcommand{\ie}{{\em i.e.\ }}

\newcommand{\al}{\alpha}

\newcommand{\del}{{\delta}}

\newcommand{\pet}{{\cal P}}

\newcommand{\bata}{\bar{\eta}}
\newcommand{\bapet}{\bar{\pet}}

\newcommand{\bett}{{\bf 1}}
\newcommand{\halv}{\frac{1}{2}}
\newcommand{\kvart}{\frac{1}{4}}
\begin{document}
\noindent
G\"{o}teborg ITP 92-48\\
October 1992\\

\vspace*{10 mm}
\begin{center}{\LARGE\bf Simple BRST quantization \\of general
gauge models.}\end{center}  \vspace*{12 mm}

\begin{center}Robert Marnelius \\
\vspace*{7 mm}
{\sl Institute of Theoretical Physics\\
Chalmers University of Technology\\
S-412 96  G\"{o}teborg, Sweden}\end{center}
\vspace*{15 mm}
\begin{abstract}
It is shown that the BRST charge $Q$ for any gauge model with a
Lie algebra symmetry may be decomposed as
$$Q=\del+\del^{\dag},\;\;\;\del^2=\del^{\dag 2}=0,\;\;\;[\del,
\del^{\dag}]_+=0$$ provided dynamical Lagrange multipliers are used but without
introducing other matter variables in $\del$ than the gauge generators in $Q$.
Furthermore, $\del$ is shown to have the form $\del=c^{\dag a}\phi_a$ (or
$\phi'_ac^{\dag a}$) where $c^a$ are anticommuting expressions in the ghosts
and Lagrange multipliers, and where the non-hermitian operators $\phi_a$
satisfy the same Lie algebra as the original gauge generators. By
means of a bigrading the BRST condition reduces to
$\del|ph\hb=\del^{\dag}|ph\hb=0$ which is naturally solved by
$c^a|ph\hb=\phi_a|ph\hb=0$ (or $c^{\dag a}|ph\hb={\phi'_a}^{\dag}|ph\hb=0$).
The
general solutions are shown to have a very simple form.\end{abstract}

\setcounter{page}{1}
\section{Introduction.}
In the general investigation performed in \cite{HM} it was noted that a BRST
quantization on an
inner product space is equivalent to a generalized Gupta-Bleuler quantization.
Furthermore, it was
shown that in this case the BRST charge may always be written in the form
\be
&&Q=\del+\del^{\dag}
\eel{e:1}
 where
\be
&&\del^2=\del^{\dag 2}=0
\eel{e:2}
and
\be
&&[\del, \del^{\dag}]_+=0
\eel{e:3}
consistent with a nilpotent $Q$. The considered physical states
\be
&&Q|ph\hb=0
\eel{e:4}
were shown to satisfy
\be
&&\del|ph\hb=\del^{\dag}|ph\hb=0
\eel{e:5}
This was performed within a particular formalism in which the ghost part of the
state space was
gauge fixed by an allowed gauge fixing \cite{Aux}. In order for \r{4} to imply
\r{5} we need a
bigrading of the state space, namely \cite{RM}
\be
&&F|k,m\hb=k|k,m\hb,\;\;\;F^{\dag}|k,m\hb=m|k,m\hb
\eel{e:6}
where $F$ satisfies
\be
&&\del=[F, Q],\;\;\;N=F-F^{\dag}
\eel{e:7}
where $N$ is the ghost number operator. The decomposition \r{1} follows then
from \r{7} since $[N,
Q]=Q$. (However, the conditions \r{2} and \r{3} restrict the possible forms of
$F$.) We have now
\be
&&Q|k,m\hb=0\; \Lra \;\del|k,m\hb=0,\;\;\del^{\dag}|k,m\hb=0
\eel{e:8}
Strictly positive normed physical states can only be obtained for states with
ghost number zero, \ie
$|m, m\hb$ for some $m$. The above properties were  summerized in section 5 in
\cite{RM}.

It is instructive to consider a typical example. Consider therefore a gauge
model with
$2m$ independent gauge generators given by $\phi_a, \phi^{\dag}_a$
($a=1,\ldots.m$) which satisfy
\be
&&[\phi_a,\phi_b]=iU_{ab}^{\;\;\;c}\phi_c\nn\\
&&[\phi_a,\phi_b^{\dag}]=0
\eel{e:9}
where $U_{ab}^{\;\;\;c}$ are constants. The BRST charge for this model may be
written in the form
\r{1} where
\be
&&\del=\phi_ac^{\dag a}-\frac{i}{2}U_{ab}^{\;\;\;c}(k_cc^{\dag a}c^{\dag
b}-c^{\dag b}c^{\dag a}k_c)
\eel{e:11}
where $c^a$ and $k_a$ are complex fermionic ghosts and their conjugate momenta
satisfying the
algebra (the nonzero part)
\be
&&[k_a, c^{\dag b}]_+=\del_{\;a}^b
\eel{e:12}
Eq.\r{11} satisfies \r{2} and \r{3}. There are two natural forms in which to
write $\del$. They are
\be
&&\del=c^{\dag a}\phi'_a
\eel{e:13}
or
\be
&&\del=\phi''_ac^{\dag a}
\eel{e:14}
where
\be
&&\phi'_a=\phi_a+\frac{i}{2}U_{ab}^{\;\;\;b}-\frac{i}{2}U_{ab}^{\;\;\;c}c^{\dag
b}k_c\nn\\
&&\phi''_a=\phi_a-\frac{i}{2}U_{ab}^{\;\;\;b}+\frac{i}{2}U_{ab}^{\;\;\;c}k_cc^{\dag b}
\eel{e:15}
Thus, $\phi''_a=\phi'_a-\frac{i}{2}U_{ab}^{\;\;\;b}$. The conditions \r{5} are
therefore naturally
solved by either
\be
&&c^a|ph\hb=0,\;\;\;\phi'_a|ph\hb=0
\eel{e:16}
or
\be
&&c^{\dag a}|ph\hb=0,\;\;\;{\phi''_a}^{\dag}|ph\hb=0
\eel{e:17}
Now consistency requires a closed algebra of the constraints. Although
\be
&&[c^a, \phi'_b]_-=[c^a, c^b]_+=0
\eel{e:18}
we have
\be
&&[\phi'_a,
\phi'_b]_-=iU_{ab}^{\;\;\;c}\phi'_c-\frac{1}{4}U_{ab}^{\;\;\;c}U_{cd}^{\;\;\;e}c^{\dag
d}k_e
\eel{e:19}
Hence, \r{16} requires also
\be
&&k_a|ph\hb=0,
\eel{e:20}
and \r{17} requires
\be
&&k^{\dag}_a|ph\hb=0,
\eel{e:21}
at least for some $a$. (One may add a term $f_{ab}^{\;\;\;c}c^{\dag b}k_c$ to
$\phi'_a$ where $f_{ab}^{\;\;\;c}$ is
symmetric in $a$ and $b$ without altering $\del$ and such that
$\phi'_a+f_{ab}^{\;\;\;c}c^{\dag b}k_c$
satisfy a closed algebra. However, $f_{ab}^{\;\;\;c}$ is then model dependent.)
 The conditions
\r{20} and \r{21} are natural additional conditions since only the states with
ghost number zero
contribute to the inner products. Eqs.\r{16} and \r{20} imply (and similarly
for \r{17} and \r{21})
\be
&&(\phi_a+\frac{i}{2}U_{ab}^{\;\;\;b})|ph\hb=0
\eel{e:22}
which is consistent. Which of the formal possibilities \r{16},\r{20} or
\r{17},\r{21} or a mixture
thereof which actually is possible depends on the explicit form of $\phi_a$ and
the state space at
hand.

In the above example as well as in all previous treatments the possibility of a
generalized
Gupta-Bleuler quantization  always restrict the possible form of the gauge
model. The gauge generators
(constraints) are always required to be possible to write as complex conjugate
pairs. In this paper
we shall show the  existence of another possibility. In fact, it will be shown
that any  bosonic
gauge model with finite number of degrees of freedom leads to a BRST charge
which may be written in
the form \r{1} where $\del$ has the simple form \r{13} or \r{14} and that the
conditions \r{5} are
naturally solved by \r{16} or \r{17}. This will be shown to be possible
provided dynamical Lagrange
multipliers are introduced. In distinction to the above example it will be
shown that $\phi'_a$ and
$\phi''_a$ may easily be made to satisfy a closed algebra so that \r{20} and
\r{21} are not necessary
to impose. Furthermore we  give general simple solutions of \r{16} and \r{17}
which reveal an
interesting general structure. (It was the proposal of \cite{MO} and the
important observations made
in the conclusions of \cite{Gen} which suggested the existence of this
construction.)

\section{Decompositions of general BRST charges.}
\setcounter{equation}{0}
Consider for simplicity  a general  gauge model with
finite number of degrees of freedom in which the gauge group is a Lie group.
Within the
Hamiltonian formulation of the corresponding BRST invariant model the BRST
charge may be
chosen to be in the general BFV form \cite{BV} ($a, b, c =1,\ldots,m<\infty$)
\be
&&Q=\psi_a\eta^a-\frac{1}{2}iU_{bc}^{\;\;a}\pet_a
\eta^b\eta^c-\frac{1}{2}iU_{ab}^{\;\;b}\eta^a + \bapet_a\pi^a
\eel{e:201}
where $\psi_a$ are the bosonic gauge generators
(constraint) satisfying
\be
&[\psi_a, \psi_b]_{-}=iU_{ab}^{\;\;c}\psi_c
\eel{e:202}
where $U_{ab}^{\;\;c}$ are the structure constants. (We consider only first
rank theories.)
 $\eta^a$ and $\bata^a$ are Faddeev-Popov (FP)
ghosts and antighosts respectively, and $\pet_a$ and $\bapet_a$
their conjugate momenta. $\pi_a$ is the conjugate momentum to the Lagrange
multiplier $v^a$.
Their fundamental algebra is (the nonzero part)
 \be
&[\eta^a, \pet_b]_+=[\bata^a, \bapet_b]_+=\del^a_{\;b},\;\;\;[\pi_b,
v^a]_-=-i\del^a_{\;b}
\eel{e:203}
which when combined with \r{202} makes $Q$ nilpotent. Since $Q$ is required to
have ghost number one,
$\eta^a$ and $\bapet_a$ have ghost number  one while $\bata^a$
and $\pet_a$ have ghost number minus one. The ghost number operator is
\be
&&N=\eta^a\pet_a-\bata_a\bapet^a
\eel{e:204}
Since $Q$ is required to be hermitian, $\psi_a$ must either be hermitian or
contain hermitian
conjugate pairs as in the example given in the introduction. Here we choose
$\psi_a$ as well as all
the other variables to be hermitian a choice one always may do.

In a first effort to decompose \r{201} according to \r{1} satisfying \r{2} and
\r{3} we introduce
the following non-hermitian  ghosts
\be
&&c^a\equiv\eta^a-i\bapet^a\nn\\
&&k_a\equiv\halv(\pet_a-i\bata_a)
\eel{e:205}
which due to \r{203} satisfy (the nonzero part)
\be
&&[k_a, c^{\dag b}]_+=\del^b_{\;a}
\eel{e:206}
In terms of these ghosts $Q$ becomes
\be
&&Q=\halv c^a(\psi_a + i\pi_a) +
\halv c^{\dag a}(\psi_a - i\pi_a)-\frac{1}{4}iU_{bc}^{\;\;a}(c^{\dag b}c^{
c}k_a+k^{\dag}_ac^{\dag b}c^{ c})
\nn\\
&&-\frac{1}{8}iU_{bc}^{\;\;a}(k^{\dag}_ac^{\dag b}c^{\dag c}+k_ac^bc^c+
k^{\dag}_ac^{ b}c^{ c}+c^{\dag b}c^{\dag c}k_a)
\eel{e:207}
Since the ghost number operator \r{204} now may be written as
\be
&&N=c^{\dag a}k_a-k^{\dag }_ac^a
\eel{e:208}
it is natural to try to define $\del$ by
\be
&&\del=[c^{\dag a}k_a, Q]
\eel{e:209}
However, although this definition together with \r{208} imply
$Q=\del+\del^{\dag}$ \r{209} is {\em
not} nilpotent when $U_{ab}^{\;\;c}\neq 0$. This problem is caused by the terms
containing
$k_ac^bc^c$ and $k^{\dag}_ac^{\dag b}c^{\dag c}$ in $Q$.
They imply
\be
&&[k^{\dag }_ac^a, \del]\neq 0
\eel{e:210}
We have therefore to get rid of these terms. For this purpose we consider a
unitary transformation
which only involve the ghosts and the Lagrange multipliers. It turns out that
what we need is a
transformation of $\bapet^a$ of the following form
\be
&&\bapet^a=M^a_{\;b}\bapet'^b+K^a_{\;b}\eta'^b
\eel{e:211}
where $M^a_{\;b}$ and $K^a_{\;b}$ are hermitian matrix elements depending on
the Lagrange multipliers
$v^a$. Hermiticity and ghost numbers are preserved by \r{211}. Since it
requires \be
&&\bata_a=(M^{-1})^b_{\;a}\bata'_b
\eel{e:212} the matrix $M^a_{\;b}$ must be nonsingular for all $v^a$. If we fix
the ghosts and
Lagrange multipliers, \ie
\be
&&\eta^a=\eta'^a,\;\;\;v^a=v'^a
\eel{e:213}
then \r{211} and \r{212} require
\be
&&\pet_a=\pet'_a-(M^{-1})^b_{\;c}K^c_{\;a}\bata'_b\nn\\
&&\pi_a=\pi'_a+iH_{ab}^{\;\;\;c}\bata'_c\eta'^b+iG_{ab}^{\;\;\;c}(\bata'_c\bapet'^b-\bapet'^b\bata'_c)
\eel{e:214}
where
\be
&&H_{ab}^{\;\;\;c}=-(M^{-1})^c_{\;d}\partial_aK^d_{\;b}\nn\\
&&G_{ab}^{\;\;\;c}=-\halv(M^{-1})^c_{\;d}\partial_aM^d_{\;b}
\eel{e:215}
The primed variables satisfy the same algebra as the original unprimed ones.
The inverse
transformation is given by \r{213} together with
\be
&&\bapet'^a=(M^{-1})^a_{\;b}(\bapet^b-K^b_{\;c}\eta^c);\;\;\;\bata'_a=M^b_{\;a}\bata_b,\;\;\;\pet'_a=\pet_a+K^b_{\;a}\bata_b\nn\\
&&\pi'_a=\pi_a+i{H'}^{\;\;\;c}_{ab}\bata_c\eta^b+i{G'}^{\;\;\;c}_{ab}(\bata_c\bapet^b-\bapet^b\bata_c)
\eel{e:216}
where
\be
&&{H'}^{\;\;\;c}_{ab}=\partial_aK^c_{\;b}-\partial_aM^c_{\;d}(M^{-1})^d_{\;e}K^c_{\;b}\nn\\
&&{G'}^{\;\;\;c}_{ab}=\halv\partial_aM^c_{\;d}(M^{-1})^d_{\;b}
\eel{e:217}

We insert now the transformation \r{211}-\r{214} into the BRST charge \r{201}.
We find then (the
primes are suppressed in the following)
\be
&&Q=\psi_a\eta^a-\halv
iU_{ab}^{\;\;\;b}\eta^a+M^a_{\;b}\pi_a\bapet^b+K^a_{\;b}\pi_a\eta^b
+iM^a_{\;b}G_{ac}^{\;\;\;c}\bapet^b+iK^a_{\;b}G_{ac}^{\;\;\;c}\eta^b-\nn\\
&&-\halv
iU_{bc}^{\;\;\;a}\pet_a\eta^b\eta^c-iT_{bc}^{\;\;\;d}\bata_d\eta^b\eta^c+
iR_{bd}^{\;\;\;c}\bapet^b\bata_c\eta^d-iV_{bd}^{\;\;\;c}\bapet^b\bapet^d\bata_c
\eel{e:218}
 where $G_{ab}^{\;\;\;c}$ is given by \r{215} and
 \be
&&T_{bc}^{\;\;\;d}\equiv -\halv
(M^{-1})^d_{\;e}\{U_{bc}^{\;\;\;a}K^e_{\;a}+K^a_{\;b}\partial_aK^e_{\;c}-
K^a_{\;c}\partial_aK^e_{\;b}\}\nn\\
&& V_{bc}^{\;\;\;d}\equiv-\halv
(M^{-1})^d_{\;e}\{M^a_{\;b}\partial_aM^e_{\;c}-
M^a_{\;c}\partial_aM^e_{\;b}\}\nn\\
&& R_{bc}^{\;\;\;d}\equiv (M^{-1})^d_{\;e}\{K^a_{\;c}\partial_aM^e_{\;b}-
M^a_{\;b}\partial_aK^e_{\;c}\}
\eel{e:219}

If we now express $Q$ in terms of non-hermitian ghosts $c^a$ and $k_a$ defined
as in \r{205} and
satisfying \r{206} we find
\be
&&Q=\halv(c^{\dag a}\psi'_a+c^a{\psi'}_a^{\dag})+\kvart(-\halv
iU_{bc}^{\;\;\;a}-iR_{bc}^{\;\;\;a}
+T_{bc}^{\;\;\;a}-V_{bc}^{\;\;\;a})c^{\dag b}c^{\dag c}k_a\nn\\
&&+\kvart(-\halv iU_{bc}^{\;\;\;a}-iR_{bc}^{\;\;\;a}
	-T_{bc}^{\;\;\;a}+V_{bc}^{\;\;\;a})k^{\dag}_ac^{ b}c^{ c}\nn\\
&&+\kvart(iU_{bc}^{\;\;\;a}+iR_{bc}^{\;\;\;a}+iR_{cb}^{\;\;\;a}
	-2T_{bc}^{\;\;\;a}-2V_{bc}^{\;\;\;a})c^{\dag b}k_ac^{ c}\nn\\
&&+\kvart(iU_{bc}^{\;\;\;a}-iR_{bc}^{\;\;\;a}-iR_{cb}^{\;\;\;a}
	+2T_{bc}^{\;\;\;a}+2V_{bc}^{\;\;\;a})c^{\dag b}k^{\dag}_ac^{ c}\nn\\
&&+\kvart(-\halv iU_{bc}^{\;\;\;a}+iR_{bc}^{\;\;\;a}
	+T_{bc}^{\;\;\;a}-V_{bc}^{\;\;\;a})k_ac^bc^c\nn\\
&&+\kvart(-\halv iU_{bc}^{\;\;\;a}+iR_{bc}^{\;\;\;a}
	-T_{bc}^{\;\;\;a}+V_{bc}^{\;\;\;a})k^{\dag}_ac^{\dag b}c^{\dag c}
\eel{e:220}
where
\be
&&\psi'_a\equiv\psi_a+K^b_{\;a}\pi_b-iM^b_{\;a}\pi_b+iK^b_{\;a}G_{bc}^{\;\;\;c}
+M^b_{\;a}G_{bc}^{\;\;\;c}+\nn\\
&&+T_{ba}^{\;\;\;b}-V_{ab}^{\;\;\;b}+
\halv iR_{ab}^{\;\;\;b}+\halv
iR_{ba}^{\;\;\;b} \eel{e:221}

In order for $\del=[c^{\dag a}k_a, Q]$ to be nilpotent the coefficients of the
terms involving
$k_ac^bc^c$ and
$k^{\dag}_ac^{\dag b}c^{\dag c}$ must vanish. This leads to the conditions
\be
&&T_{bc}^{\;\;\;a}-V_{bc}^{\;\;\;a}=0\nn\\
&&R_{bc}^{\;\;\;a}-R_{cb}^{\;\;\;a}-U_{bc}^{\;\;\;a}=0
\eel{e:222}
which may be simplified to
\be
&&L^c_{\;a}\partial_cL^d_{\;b}-L^c_{\;b}\partial_cL^d_{\;a}=-iU_{ab}^{\;\;\;c}L^d_{\;c}
\eel{e:223}
where we have introduced the complex matrix
\be
&&L^b_{\;a}=M^b_{\;a}+iK^b_{\;a}
\eel{e:224}
The meaning of the conditions \r{223} may be obtained by considering the
$\pi$-term in the
expression \r{221} of $\psi'_a$ and ${\psi'}_a^{\dag}$. They are
\be
&&\phi_a\equiv -iL^b_{\;a}\pi_b,\;\;\;\phi'_a\equiv i{L^{\dag}}^{ b}_{\;a}\pi_b
\eel{e:225}
Conditions \r{223} imply then
\be
&&[\phi_a, \phi_b]=iU_{ab}^{\;\;\;c}\phi_c\nn\\
&&[\phi'_a, \phi'_b]=iU_{ab}^{\;\;\;c}\phi'_c
\eel{e:226}
The Jacobi identities
\be
&&\sum_{(abc)}[[\phi_a, \phi'_b], \phi_c]=0
\eel{e:227}
require furthermore
\be
&&[\phi_a, \phi'_b]=0
\eel{e:228}
or equivalently
\be
&&L^c_{\;a}\partial_c{L^{\dag}}^{d}_{\;b}-{L^{\dag}}^{
c}_{\;b}\partial_cL^d_{\;a}=0
\eel{e:229}
$L^c_{\;a}$ may then be recognized as the left vielbein field on the group
manifold with imaginary
group coordinates $\theta^a=iv^a$. $\phi_a$ and $\phi'_a$ in \r{225} may be
interpreted as
generators of left and right translations on the group manifold. One may use
\r{223} and \r{229} to
solve for $L^a_{\;b}$ assuming that $L^a_{\;b}$ is analytic in $v^a$ and that
$L^a_{\;b}(v^a=0)=\del^a_{\;b}$. Since
\be
&&M^a_{\;b}=\halv(L^a_{\;b}+L^{\dag a}_{\;b})\nn\\
&&K^a_{\;b}=-\halv i(L^a_{\;b}-L^{\dag a}_{\;b})
\eel{e:230}
one may notice that $M^a_{\;b}(-v)=M^a_{\;b}(v)$, $M^a_{\;b}(v=0)=\del^a_{\;b}$
and
$K^a_{\;b}(-v)=-K^a_{\;b}(v)$.

The conditions \r{223} and \r{229} reduce now the BRST charge \r{220} to the
following form
\be
&&Q=\halv c^{\dag a}\Phi_a+\halv\Phi^{\dag}_ac^a
\eel{e:231}
where
\be
&&\Phi_a=\psi_a-iL^b_{\;a}\pi_b-\halv\partial_bL^{
b}_{\;a}+g_{ab}^{\dag\;\;b}+2g_{ab}^{\;\;\;c}k_c^{\dag}c^b-\halv
iU_{ab}^{\;\;\;c}c^{\dag b}k_c
\eel{e:232} where in turn
\be
&&g_{ab}^{\;\;\;c}\equiv\kvart iU_{ab}^{\;\;\;d}L^e_{\;d}(M^{-1})^c_{\;e}
\eel{e:233}
Notice that
\be
&&\del\equiv[c^{\dag a}k_a, Q]=\halv c^{\dag a}\Phi_a
\eel{e:234}
Since (cf \r{19}!)
\be
&&[\Phi_a, \Phi_b]=iU_{ab}^{\;\;\;c}\Phi_c-\kvart
U_{ab}^{\;\;\;c}U_{ce}^{\;\;\;d}c^{\dag e}k_d\nn\\
&&[\Phi_a, c^{\dag b}]=-\halv iU_{ac}^{\;\;\;b}c^{\dag c}
\eel{e:235}
one easily verifies that $\del^2=0$. $[\del, \del^{\dag}]_+=0$ follows then
automatically since
$Q^2=0$ by construction. (However, we have also checked it explicitly.) Thus,
we have achieved our
goal to prove the existence of the decomposition \r{1} with the properties
\r{2} and \r{3} for a
general gauge model.

\section{The implied general Gupta-Bleuler quantization.}
\setcounter{equation}{0}
If we impose the bigrading considered in the introduction then we may solve
$Q|ph\hb=0$ by
\be
&&\del|ph\hb=0,\;\;\;\del^{\dag}|ph\hb=0
\eel{e:301}
In view of the general result that $\del=\halv c^{\dag a}\Phi_a$ these
condition are naturally solved
by the conditions
\be
&&c^a|ph\hb=0,\;\;\;\Phi_a|ph\hb=0
\eel{e:302}
However, the algebra in \r{235} requires
\be
&&k_a|ph\hb=0
\eel{e:303}
at least for some $a$. Although this condition only eliminates solutions with
zero inner products,
it may be avoided if we change the last conditions in \r{302}. We notice then
that $\del$ is
unaffected if we replace $\Phi_a$ by $\Phi_a+f_{ab}^{\;\;\;c}c^{\dag b}k_c$
where $f_{ab}^{\;\;\;c}$
is symmetric in $a$ and $b$. In particular we may choose
 \be
&&\Phi'_a\equiv\Phi_a+\halv(L^d_{\;a}\partial_dL^e_{\;b}+
L^d_{\;b}\partial_dL^e_{\;a})(L^{-1})^c_{\;e}c^{\dag
b}k_c
\eel{e:304}
which satisfies
\be
&&[\Phi'_a, \Phi'_b]=iU_{ab}^{\;\;\;c}\Phi'_c
\eel{e:305}
Since
\be
&&[\Phi'_a, c^b]=-2g_{ac}^{\;\;\;b}c^c
\eel{e:306}
we may then consistently solve \r{301} by
\be
&&c^a|ph\hb=0,\;\;\;\Phi'_a|ph\hb=0
\eel{e:307}
Another possibility is
\be
&&c^{\dag a}|ph\hb=0,\;\;\;{\Phi''_a}^{\dag}|ph\hb=0
\eel{e:308}
where $\Phi''_a$ is defined by $\del=\halv\Phi''_ac^{\dag a}$ and which may be
chosen to be
\be
&&\Phi''_a=\Phi'_a-\partial_bL^b_{\;a}
\eel{e:309}
which also satisfies the closed algebra \r{305}.

\section{Solving general Gupta-Bleuler quantization.}
\setcounter{equation}{0}
Consider the general BRST charge \r{201} in terms of the original ghost
variables. Consider then
in particular the BRST invariant operator
\be
&&A\equiv[\rho, Q]_+
\eel{e:401}
where
\be
&&\rho\equiv v^a\pet_a
\eel{e:402}
It is explicitly
\be
&&A\equiv(\psi_a+\psi_a^{gh}) v^a-i\pet_a\bapet^a
\eel{e:403}
where
\be
&&\psi_a^{gh}\equiv \halv i U_{ab}^{\;\;\;c}(\pet_c\eta^b-\eta^b\pet_c)
\eel{e:404}
is hermitian and satisfies the same algebra as $\psi_a$. We notice then that
\be
&&e^A\eta^ae^{-A}=(L^{\dag -1})^a_{\;c}(L^c_{\;b}\eta^b-i\bapet^c)\nn\\
&&e^{-A}\eta^ae^{A}=(L^{ -1})^a_{\;c}({L^{\dag}}^{ c}_{\;b}\eta^b+i\bapet^c)
\eel{e:405}
These expressions may be compared to $c^a$ and $c^{\dag a}$ in terms of the
original ghost variables
(use \r{216}):
\be
&&c^a=\eta'^a-i\bapet'^a=(M^{-1})^a_{\;b}(L^b_{\;c}\eta^c-i\bapet^b)
\eel{e:406}
Obviously
\be
&&c^a|ph\hb=0\;\Lra\;e^A\eta^ae^{-A}|ph\hb=0
\eel{e:407}
So far nothing is achieved. However, consider now the same transformations  of
the conjugate
momentum to $v^a$. We find
\be
&&e^A\phi_ae^{-A}=\psi_a+\psi_a^{gh}-iL^b_{\;a}\pi_b-
i(L^{-1})^c_{\;e}\partial_dL^e_{\;a}\pet_c\bapet^d
 \eel{e:409}
where
\be
&&\phi_a=-iL^b_{\;a}\pi_b
\eel{e:410}
Expressing the right hand side of \r{409} in terms of the new ghost variables
by means of
\r{211}-\r{214} we find
 \be
&&e^A\phi_ae^{-A}=\Phi'_a+({L^{\dag}}^{d}_{\;b}\partial_dL^e_{\;a}(L^{-1})^c_{\;e}-g_{ab}^{\dag\;c})k_cc^b
\eel{e:411}
Hence, $c^a|ph\hb=0,\;\Phi'_a|ph\hb=0$ has the solution
\be
&&|ph\hb=e^A|\phi\hb
\eel{e:412}
where $|\phi\hb$ satisfies
\be
&&\eta^a|\phi\hb=\pi_a|\phi\hb=0
\eel{e:413}
which is trivially solved. Similarly
\be
&&e^{-A}\phi'_ae^{A}=(e^A\phi_ae^{-A})^{\dag}-\partial_bL^{\dag
b}_{\;a}=\nn\\
&&=(\Phi''_a)^{\dag}-(L^{
d}_{\;b}\partial_d{L^{\dag}}^{ e}_{\;a}(L^{\dag
-1})^c_{\;e}-g_{ab}^{\;\;c})k^{\dag}_cc^{\dag b}
\eel{e:414}
where
\be
&&\phi'_a=i{L^{\dag}}^{ b}_{\;a}\pi_b
\eel{e:415}
Hence,
\be
&&c^{\dag a}|ph\hb=0,\;{\Phi''_a}^{\dag}|ph\hb=0
\eel{e:416}
is solved by
\be
&&|ph\hb=e^{-A}|\phi\hb
\eel{e:417}
where $|\phi\hb$ satisfies \r{413}. Since $e^{\pm A}$ is BRST invariant also
$|\phi\hb$ must be BRST
invariant. In fact, $Q|\phi\hb=0$ follows from \r{413}. Notice that $|\phi\hb$
does not belong to an
inner product space although $|ph\hb$ must do.

\section{Further properties of the general solutions.}
\setcounter{equation}{0}
We have found that the general Gupta-Bleuler quantization implied by
$\del|ph\hb=\del^{\dag}|ph\hb=0$ lead to two different solutions
\be
&&|ph\hb_{\pm}=e^{\pm[\rho, Q]}|\phi\hb
\eel{e:501}
We must therefore require that these two sets of solutions yield equivalent
physical results.
 Now, there are even more solutions. In fact
\be
&&|ph\hb_{\al}=e^{\al[\rho, Q]}|\phi\hb
\eel{e:502}
is  a satisfactory solution for any real $\al\neq0$. The reason for this is
that the BRST charge
$Q$ is invariant under the unitary transformation
\be
&&(\pi_a, v^a) \;\rightarrow\; (\frac{\pi_a}{\al}, \al v^a),\;\;\;(\bapet^a,
\bata_a) \;\rightarrow\;
(\al\bapet^a, \frac{\bata_a}{\al})
\eel{e:503}
where $\al$ is a real positive constant. However, $\del$ and $\rho$ are not
invariant under \r{503}.
Instead we have ($\rho=\pet_av^a$)
\be
&&\rho\;\rightarrow\;\al\rho,\;\;\;\del\;\rightarrow\;\del'=U\del
U^{\dag},\;\;|ph\hb_{\al}=U|ph\hb_{\pm}
 \eel{e:504}
where $e^{\al[\rho, Q]}=Ue^{\pm[\rho, Q]}U^{\dag}$. Notice that
$|\phi\hb'=U|\phi\hb$
satisfies \r{413} for any $\al\neq0$
in \r{502}. Thus, we have proved that \r{502} reduces to the two possible
solutions \r{501}. (Notice
that the solution \r{502} corresponds to a different bigrading than \r{501}.)

Consider now the inner product of \r{502}.  We find
\be
&&\vb\phi|e^{2\al[\rho, Q]}|\phi\hb=\,'\vb\phi|e^{\pm[\rho, Q]}|\phi\hb'
\eel{e:505}
where we have  performed a unitary  scale transformation of the type \r{503}.
Since only the ghost
number zero part of $|\phi\hb'$ contribute in the inner product we have
\be
&&_{\al}\vb ph|ph\hb_{\al}=\,'\vb\phi|e^{\pm[\rho,
Q]}|\phi\hb'=\,_0'\vb\phi|e^{\pm[\rho,
Q]}|\phi\hb'_0 \eel{e:507}
where $|\phi\hb'_0$ satisfies
\be
&&\bata_a|\phi\hb'_0=0
\eel{e:508}
apart from \r{413}. (Notice that $[\rho, Q]$ has ghost number zero.)
Eq.\r{508} implies then that
\be
&&|ph\hb_{\pm}=e^{\pm[\rho, Q]}|\phi\hb'_0
\eel{e:509}
satisfies
\be
&&k_a|ph\hb_+=0,\;\;\;k_a^{\dag}|ph\hb_-=0
\eel{e:510}
apart from \r{307} and \r{308}. These conditions imply that the solutions
\r{509} have ghost number
zero. Eq.\r{510}  could also follow from antiBRST invariance. (Notice, however,
that $[\rho,
Q]$ is only antiBRST invariant for totally antisymmetric structure constants
$U_{abc}$ \cite{Hw}.) In
$|\phi\hb'_0$ in \r{509} the ghost as well as  the Lagrange multiplier
dependence is completely
fixed. This means that all dependence on ghosts and Lagrange multipliers in
\r{509} is through the
factor $e^{\pm[\rho, Q]}$. In the inner products all ghost dependence
disappears after a reduction of
the factor $e^{\pm[\rho, Q]}$ and only an integration over the Lagrange
multiplier $v^a$ remains.
However, this is the tricky part. Firstly, there is an ambiguity in how the
Lagrange multipliers
should be quantized. In \cite{Gen} the following general rule was extracted
from \cite{MO}: The
Lagrange multipliers must be quantized with opposite metric states to the
variables which the
corresponding gauge generators eliminate. Thus, one has to specify exactly how
the original matter
variables should be quantized. Secondly, one has to specify the range of the
Lagrange multipliers.
Typically one may find that the range must be restricted in order for $|ph\hb$
to belong to an inner
product space. In such a case they should probably at most be restricted to the
group manifold. In
the case when the matter variables are quantized with positive metric states we
find formally  \be
&&_{\al}\vb ph|ph\hb_{\al}\propto\int\! dx \del(\psi_a')|\phi(x)|^2
\eel{e:511}
since when $v^a$ is quantized with negative metric states the spectral
representation
requires imaginary eigenvalues \cite{Gen}, which means that we obtain
\be
&&\int\! d^mv e^{\pm iv^a\psi_a'}\propto\del(\psi'_a)
\eel{e:512}
in \r{507}.  We have also introduced a complete set of states $|x\hb$ in the
matter space in \r{511},
\ie $\int\! dx|x\hb\vb x|=\bett$.

\section{Final remarks.}
\setcounter{equation}{0}
We have proved that the BRST charge $Q$ may be decomposed as
\be
&&Q=\del+\del^{\dag}
\eel{e:601}
where $\del$ satisfies \r{2} and \r{3} for any gauge model with finite number
of degrees of freedom
and whose gauge symmetry is a Lie group. Furthermore, we have shown that the
general solutions are
very simple to \be
&&\del|ph\hb=\del^{\dag}|ph\hb=0,
\eel{e:602}
which is equivalent to $Q|ph\hb=0$ when the bigrading \r{6} is used. However,
the
solutions could not be made completely explicit since it remains to specify the
properties of the
Lagrange multipliers which depend on the more detailed structure of the matter
states. Each model
has therefore to be investigated separately in order to determine these
properties exactly. Some
examples are given in \cite{Proper}.

The general decomposition \r{601} found in section 2 and defined up to the
scale transformations
\r{504} are not the only possible ones. However, other decompositions will
involve other matter
variables than the gauge generators $\psi_a$ which will make $\del$ less
attractive and the
corresponding solutions not so useful as those found in this paper. Now we have
only proved \r{601}
for bosonic gauge models with finite number of degrees of freedom. However, we
believe that the
generalization to graded gauge groups is straight-forward and that most
formulas are trivially
generalized to this case. Examples of this type are given in \cite{Proper}.
Also theories with
second class constraints or anomalous gauge theories should be possible to
treat by this method
since the bigrading leading to \r{602} were shown to be essential for such
models in \cite{RM}.
However, this application involves additional features since the property \r{3}
is no longer
valid. The generalization to gauge theories of higher rank than the first is of
course highly
nontrivial and to which the present method is not immediately generalizable.
(However, one could
try transformations of the type \r{405}.)  The method is also not trivially
generalizable to gauge
models with infinite degrees of freedom. However, for strings the oscillator
part may be
decomposed like in the introduction as was first made by Banks and Peskin
\cite{BP}. Our method
suggests then that if one also introduces dynamical Lagrange multiplier fields
also the zero mode
part will be possible to decompose as in \r{601} making the complete BRST
charge in the form
\r{601}.

\end{document}